\documentclass[12pt]{article}
\usepackage{latexsym}
\usepackage{amsmath}
\usepackage[dvips]{graphicx}
\usepackage{textcomp}

\newcommand{\be}{\begin{equation}}
\newcommand{\ee}{\end{equation}}
\newcommand{\ba}{\begin{array}}
\newcommand{\ea}{\end{array}}

\author{Fabio Cardone $^{1}$, Roberto Mignani $^{2,3,*}$,
Andrea Petrucci $^{1}$\\ \\ $^{1}$Istituto per lo Studio dei
Materiali Nanostrutturati (ISMN — CNR) \\ Via dei Taurini - 00185
Roma, Italy
  \\
$^{2}$Dipartimento di Fisica "E.Amaldi", Universit\`a degli Studi
"Roma Tre" \\ Via della Vasca Navale, 84 - 00146 Roma, Italy \\$^{3}$INFN Sezione di Roma Tre, Italy  \\
* Corresponding author: mignani@fis.uniroma3.it}
\date{}
\title{Reply to "Comment on 'Piezonuclear decay of thorium'
[Phys. Lett. A 373 (2009) 1956]" by L. Kowalski }
\begin{document}
\maketitle \abstract{In a paper appearing in this issue of Physics
Letters A, Kowalski raises some critical comments on the experiment
[F. Cardone, R. Mignani, A. Petrucci, Phys. Lett. A 373 (2009) 1956]
that we carried out by cavitating a solution of thorium-228. The
experiment highlighted the anomalous decay of thorium, thus
confirming the results previously obtained by Urutskoev et al. by
explosion of titanium foils in solutions. In this Letter, we reply
to these comments. We agree with Kowalski that critical comments are
one of the key factors of the process to improve the quality of
experiments and the interpretation of results. However we do hope
that these comments together with the details provided in the
replies will promote further and better experiments which are
certainly worth performing in order to shed a brighter light on this
issue, as Kowalski himself suggests in his comment.}

\section{Reply to comments}

In the article 'Comments on "Piezonuclear decay of thorium" appeared
on this same issue of Physics Letters A ~\cite{kowa} Kowalski puts
forward some remarks and suggestions about our work on the possible
effects of cavitation on Thorium-228 ~\cite{torio}. The author
highlights what he thinks are the possible shortcomings of the paper
~\cite{torio} and gives us the chance to provide some more
clarifications. We will follow his manuscript and try to give an
answer or more details as we encounter his remarks. However, first
of all we want to complain about the fact that, after reading a few
times the text of the comment, it looks like that the author
completely ignored our reply ~\cite{reply} to the comment by
Ericsson et al. ~\cite{swedish} where he would have found out more
information and suggestions that would have helped him in
interpreting the results in a more comprehensive way. In the
introduction the author mentions the fact that Marie Curie and her
contemporaries tried to speed up alpha emission by different types
of effects without achieving any success. The purpose of the
experiments that we have conducted so far with ultrasounds,
cavitation and shockwaves~\cite{piezoneutr, neutropiezo, trasmut,
eur1,eur2} has never been devoted to speed up alpha emission,
although the experiment with Thorium might seem to be aimed at this
target. It has turned out that cavitation is able to generate
locally very peculiar conditions of pressure, temperature and energy
density that bring about new phenomena like, for instance,
sonoluminescence and other anomalous effects which seem to involve
atomic species at their nuclear level ~\cite{storms}. Although some
researchers frown at these phenomena due to, they say, the lack of
convincing proofs, we think that they are worth of being studied as
they are likely to be the tip of the iceberg of a completely new
physics. The decision of subjecting to cavitation a solution of
Thorium was based on three experimental evidences: the anomalous
transmutations that we detected in solutions subjected to cavitation
~\cite{trasmut, eur1,eur2}, the emission of bursts of neutrons
detected from iron solutions treated by cavitation
~\cite{piezoneutr, neutropiezo} and the evidences achieved by
Russian teams ~\cite{uruttrans, volktor} of the modified secular
equilibrium of Thorium in solutions subjected to electric explosions
of Titanium thin wires. Our target, in carrying out these
experiments with Thorium, was only to check whether cavitation, that
brings about anomalous transmutations and emission of neutrons from
solution of stable elements, might affect Thorium as well and
produce effects similar to those induced by electric explosions of
the Russian teams. Moreover, we would like to add that according to
some predictions of our theory ~\cite{energeo, defspti} all of these
anomalous effects (anomalous transmutations, emission of neutrons
from Iron, and all of the evidences collected in 20 years of low
energy nuclear reactions ~\cite{storms}) need very precise local
condition to take place, in terms of energy density and time of
release of this energy, which are not easy at all to be
simultaneously achieved. From this perspective, the attempts to
speed up alpha emission carried out between the end of the XIX and
the beginning of the XX century were very unlikely to be
systematically successful because of the difficulty to make out
these effects among other more evident phenomena. The first and main
remark of Kowalski is about the reduction of the concentration of
Thorium and the chance that this reduction might be due to a
redistribution of the atoms of Thorium rather than its
transmutation. In particular he suggests that "\emph{Cavitation
generates reactive chemical species like ozone or hydrogen peroxide.
This leads to progressive accumulation of thorium on available
surfaces, such as inner walls of the glass container, or the
cavitator. If this were to happen then the solution removed from the
vessel, after the treatment, would indeed be less concentrated than
the solution before the treatment}".

The hypothesis that the Thorium was deposited on the walls on the
vessel is legitimate and indeed we thought of it. However, the
convective motions of the liquid in the vessel induced by
ultrasounds formed a continuous flow all along the walls and the
surface of the sonotrode that prevented the atomic species from
being adsorbed on them\footnote{In our previous works
~\cite{piezoneutr,neutropiezo} we specified that the vessel was
suitably chosen to be a true reaction chamber that would contribute
to the generate the cavitation in its central part between the
sonotrode tip and the bottom of the chamber and would contribute to
produce convective motions along the lateral walls and the bottom.
Vessels of different shape and dimension were tested and examined
for the purpose.}.This consideration, supported by the evidences of
these convective flows, convinced us that the hypothesis of
adsorption could be put off to future and more accurate
investigations. The author complains about the lack of information
on the etching procedure of CR39. They were etched by a solution of
NaOH 6.5N at a temperature of 90 degrees centigrade for 4.5 hours.
In replying to this request we refer, as well, to the last statement
of the author that complains about the sentence in our manuscript
~\cite{torio} in which we talked about CR39 and alpha particles and
said "a polycarbonate whose energy calibration is just designed to
detect those emitted by Radon-222". This sentence, which is indeed
too short, and hence generated the remark, has to be interpreted by
referring the calibration not to the CR39, but to the etching
procedure which was suggested by the manufacturer in order to obtain
tracks comparable to those on the CR39 chips provided as calibration
reference and irradiated by Radon-222. In a further remark, the
author complains about the lack of information on the measured track
densities. We inspected the CR39 chips by an optical microscope, but
concentrated on the "star-shaped" tracks only in order to establish
that their shape and features were all compatible on the different
chips. The decision to pay attention only to the "star-shaped"
tracks finds its reason in the details provided in order to answer
another remark of Kowalski. This remark is about these "star-shaped"
tracks on the CR39 which are brought about by 5 alpha particles
emitted during the decay chain of Thorium-228 down to Lead-208.

Kowalski says that in a agitated liquid containing Thorium, unlike
in a solid, the tracks on the CR39 are not likely to be
"star-shaped" because, due to the shaking of the solution, a nucleus
of Thorium would be separated from its daughter nuclides and hence
the emitted five alpha particles from Thorium-228 and its daughters
down to the stable Lead-208 would not hit the CR39 on the same spot.
This remark refers to an issue which was considered at length before
defining the most suitable procedure to detect unmistakably by CR39
the tracks of Thorium. We knew that on the CR39 surface, after
etching, there would be both tracks due to alpha particles of the
natural background radioactivity and the tracks due to the alpha
particles of Thorium. Although 0.03 ppb of Thorium in 250 ml of
deionised bidistilled water mean an activity of about 10$^{-6}$ Ci
which is 3 to 5 orders of magnitude higher than the natural
radioactivity of water, one has to consider that the range of alpha
particle of 5 MeV in water is about 40 microns. This circumstance
radically reduces the volume of the solution that surrounds the
upper free side of the CR39\footnote{We remind that the CR39 chip
was inside the vessel, laid horizontally on its bottom and aligned
with the sonotrode tip.}  and hence reduces the number of Thorium
atoms which can produce tracks. All considered and estimated, we
understood that, had we based our comparisons among the differently
treated CR39 chips on the density of tracks, it might have been hard
to achieve a sufficiently clear picture.

Thus we decided to base our measurement on the number of the
unmistakable star-shaped tracks and hence designed a measurement
protocol that allowed the formation of these peculiar tracks. We
will briefly illustrate this protocol here. Let's focus our
attention on the CR39 chip in the vessel with non cavitated solution
and on the CR39 chip in the vessel with cavitated solution. The two
vessels were prepared at the same time and as soon as they were
ready we began the cavitation of one of them. The cavitation lasted
90 minutes and the two vessels were kept in two separate rooms. In
the non cavitated vessel the solution was, of course, still for 90
minutes. At the end these 90 minutes we turned off the ultrasounds
and took out the CR39 chip from the non cavitated solution being
careful to leave a layer of solution on the upper side of it. We let
this layer evaporate. As to the CR39 chip in the cavitated solution,
we did not take it out at the end of the 90 minutes but waited for
further 90 minutes since we wanted the solution to cool and be still
as it had been still the non cavitated one in order to let the heavy
Thorium atoms to deposit as it must have happened in the other
solution. After these further 90 minutes we took out the CR39 chip
from the solution as we had done for the non cavitated sample and
let the layer evaporate. As the liquid evaporated, some atoms of
Thorium deposited on the upper surface of the CR39 and together with
its daughter nuclides produced on the same spot the "star-shaped"
tracks. The last remarks of Kowalski is about the activity of the
solutions that we used and the lack of information about the special
equipment for handling highly radioactive liquid sources. He
estimated that the activity of our solutions was 1 Ci per litre.
This is a huge activity indeed, but certainly not the activity of
our solutions which was between $10^{-5}$ and $10^{-6}$ Ci for 250
ml (calculated and measured). We would like to point out that this
value of activity is compatible with that of the radioactive sources
which do not require the Nuclear Regulatory Commission (NRC) licence
to be handled and can be sold to the general public. We reckon that
the author must have misinterpreted the concentration of Thorium
which is 0.03 ppb (part per billion) that is 0.03 micro grams of
Thorium in 1 litre of solution. As to the precautions adopted to
work with these solutions, we certainly complied with all the
security measures, but we reckon that this information is of no use
in a scientific letter. Nevertheless, we report here that these
experiments were conducted in collaboration with the Italian Army at
the Nuclear-Chemical-Bacteriological facilities. We planned and
supervised the experiments and interpreted its results, but most of
the operations were carried out by authorised and trained personnel
in a hot cells.

\section{Conclusions}

We certainly agree with the author that critical comments contribute
to improve the quality of further experiments and their
interpretations. However, together with him we reckon that "they are
no substitute for additional experiments". Even if he were convinced
by the arguments that we presented in ~\cite{torio} and by the
convergence of our results ~\cite{torio} with those obtained by
Russian teams ~\cite{uruttrans,volktor}, further and better
experiments must be performed. We do hope that these comments and
replies with additional details will encourage other research teams
to experiment and not just comment on this issue in order to obtain
further indications either in agreement or in disproof.


\begin{thebibliography}{90}
\bibitem{kowa} L. Kowalski, Phys. Lett. A (this issue).

\bibitem{torio} F. Cardone, R. Mignani, A. Petrucci, Phys. Lett. A 373 (2009) 1956-1958.

\bibitem{reply} F. Cardone, R. Mignani, A. Petrucci, Phys. Lett. A 373 (2009) 3797-3800.

\bibitem{swedish}G. Ericsson, S. Pomp, H. Sjöstrand, E. Traneus, Phys. Lett. A 373 (2009) 3795-3796.


\bibitem{piezoneutr}F. Cardone, G. Cherubini, A. Petrucci, Phys. Lett. A 373 (2009) 862

\bibitem{neutropiezo}F. Cardone , G. Cherubini, R. Mignani, W. Perconti, A. Petrucci, F. Rosetto, G. Spera, Ann. Fond. L. de Broglie, (in press), arXiv:0710.5115.

\bibitem{trasmut} F. Cardone, R. Mignani, Int. J. Mod. Phys. B 17, 307 (2003).

\bibitem{eur1}F. Cardone, R. Mignani, W. Perconti, E. Pessa, G. Spera, Jour. Radioanalytical Nucl. Chem. 265, 151 (2005).

\bibitem{eur2}F. Cardone, R. Mignani, W. Perconti, E. Pessa, G. Spera, Gravitation and Cosmology 11, 1-2 (41-42) (2005).

\bibitem{storms}E. Storms, The Science of Low Energy Nuclear Reaction - A Comprehensive Compilation of Evidence and Explanations about Cold Fusion. World Scientific, Singapore (2007).

\bibitem{uruttrans} L.I. Urutskoev, V.I. Liksonov, V.G. Tsinoev, Ann. Fond. L. de Broglie 27, 701 (2002).

\bibitem{volktor}A.G. Volkovish, A.P. Govorum, A.A. Gulyaev, S.V. Zhukov, V.L. Kuznetsov, A.A. Rukhadze, A.V. Steblevskii, L.L. Urutskoev, Ann. Fond. L. de Broglie, 30, 1, (2005).

\bibitem{energeo} F. Cardone, R. Mignani, Energy and Geometry, World Scientific, Singapore, (2004).

\bibitem{defspti}F. Cardone, R. Mignani, Deformed Spacetime, Springer, Heidelberg-Dordrecht, (2007).






\end{thebibliography}
\end{document}